\documentclass[%
 aip,
 jmp,%
 amsmath,amssymb,
 reprint,%
nobalancelastpage
]{revtex4-1}

\usepackage{graphicx}
\usepackage{dcolumn}
\usepackage{bm}
\usepackage{hyperref} 
\usepackage{color}
\usepackage{multirow,makecell,booktabs}
\usepackage{xcolor}
\usepackage{enumitem}

\begin{document}


\title{Operation of Cs-Sb-O activated GaAs in a high voltage DC electron gun at high average current
}

\author{Jai Kwan Bae}
\email{jb2483@cornell.edu}
\author{Matthew Andorf}
\affiliation{Cornell Laboratory for Accelerator-Based Sciences and Education, Cornell University, Ithaca, NY, USA}
\author{Adam Bartnik}
\affiliation{Cornell Laboratory for Accelerator-Based Sciences and Education, Cornell University, Ithaca, NY, USA}
\author{Alice Galdi}
\affiliation{Department of Industrial Engineering, University of Salerno, Fisciano (SA), Italy}

\author{Luca Cultrera}
\affiliation{Brookhaven National Laboratory, Upton, NY, USA}

\author{Jared Maxson}
\author{Ivan Bazarov}

\affiliation{Cornell Laboratory for Accelerator-Based Sciences and Education, Cornell University, Ithaca, NY, USA}

\date{\today}

\begin{abstract}

Negative Electron Affinity (NEA) activated GaAs photocathodes are the most popular option for generating a high current ($>$ 1 mA) spin-polarized electron beam. Despite its popularity, a short operational lifetime is the main drawback of this material. Recent works have shown that the lifetime can be improved by using a robust Cs-Sb-O NEA layer with minimal adverse effects. In this work, we operate GaAs photocathodes with this new activation method in a high voltage environment to extract a high current.
We observed spectral dependence on the lifetime improvement.
In particular, we saw a 45\% increase in the lifetime at 780 nm for Cs-Sb-O activated GaAs compared to Cs-O activated GaAs.

\end{abstract}

\maketitle

\section{Introduction}

An electron source capable of sustaining high-intensity production of spin-polarized electrons is highly desired for nuclear and high energy physics applications.\cite{cardman2018,adolphson2011large}
Furthermore, electron microscopy technologies can utilize spin-polarized electrons to study magnetization in materials and nanostructures.\cite{suzuki2010_RealTimeMagnetic,kuwahara2012_30kVSpinpolarizedTransmission,vollmer2003}
A Negative Electron Affinity (NEA) activated GaAs in a high voltage electron gun is the only viable option to produce a highly spin-polarized electron beam at a high current.
However, the extreme vacuum sensitivity of the NEA layer is the main drawback that results in rapid Quantum Efficiency (QE) degradation over time; hence it has a short operational lifetime compared to other types of photocathodes.\cite{grames2011_ChargeFluenceLifetime, bae2018_RuggedSpinpolarizedElectron,bae2020_ImprovedLifetimeHigh,cultrera2020_LongLifetimePolarized}


Quantum mechanical selection rules in optically stimulated electronic transition must be leveraged to achieve the photoemission of spin-polarized electrons from bulk GaAs.\cite{pierce1976,liu2017_ComprehensiveEvaluationFactors} Using circularly polarized light with photon energy slightly larger than the bandgap, electrons can be excited from the top of the valence band to the bottom of the conduction band with polarization up to a theoretical limit of 50\%.\cite{pierce1976} Extracting these electrons from the photocathode requires a vacuum interface with NEA condition.
As in bulk GaAs, most layered GaAs-based photocathodes can be designed to achieve a sufficiently large QE and high electron spin polarization when excited with wavelengths close to 780 nm.\cite{liu2016_RecordlevelQuantumEfficiency}
Therefore, the operational lifetime at 780 nm is crucial for large accelerator facilities applications.

Three main mechanisms are responsible for the rapid degradation of QE: ion back-bombardment,\cite{grames2011_ChargeFluenceLifetime,cultrera2011_PhotocathodeBehaviorHigh,liu2016_EffectsIonBombardmenta} chemical poisoning,\cite{chanlek2014_DegradationQuantumEfficiency} and thermal desorption of the activation layer.\cite{kuriki2011_DarklifetimeDegradationGaAs}
Ion back-bombardment occurs when the electron beam ionizes the residual gases in the vacuum. The positively charged ions are accelerated towards the cathode and degrade the cathode performance.\cite{grames2011_ChargeFluenceLifetime,cultrera2011_PhotocathodeBehaviorHigh}
Chemical poisoning of the cathode happens as the NEA activation layer chemically reacts with the residual gases in the vacuum.\cite{chanlek2014_DegradationQuantumEfficiency}
As ion back-bombardment and chemical poisoning degrade the QE during beam operation, a higher laser intensity is required to maintain the beam current at the same level, and the increased laser power might further accelerate the thermal desorption of the NEA layer.
Among the three mechanisms, the ion back-bombardment is considered the most significant contributor to performance degradation during high current operations. It has been shown that such a process can be minimized by extracting electrons a few mm off from the electrostatic center, limiting the photocathode active area, increasing the laser spot size, and applying a positive bias to the anode.\cite{grames2011_ChargeFluenceLifetime,yoskowitz2021_IMPROVINGOPERATIONALLIFETIME}

Ion back bombardment can degrade QE by three different processes.
High energy ions ($\gtrsim$ 5 keV) can interact with lattice atoms and create vacancies, directly damaging the GaAs structure.\cite{biswas2019_StudyPhotocathodeSurfacea,liu2016_EffectsIonBombardmenta}
Lower energy ions ($\lesssim$ 500 eV) can penetrate the cathode surface and create interstitial defects near the vacuum interface. These defects will affect the diffusion length of electrons and hence lower the escape probability of photoelectrons.\cite{liu2016_EffectsIonBombardmenta}
The last possibility is that ions between 500 eV and 5 keV can interact with atoms on the surface and remove the NEA activation layer via sputtering.\cite{biswas2019_StudyPhotocathodeSurfacea}
The QE degradation is expected to be a convolution of these three processes, but the relative contribution of each is not well understood and left to be explored.
Understanding the convolution would be critical to potential mitigation.

Traditional NEA activation layers such as Cs-O$_2$ and Cs-NF$_3$ form a submonolayer that is chemically reactive and weakly bound to the surface.\cite{kuriki2011_DarklifetimeDegradationGaAs,chanlek2014_DegradationQuantumEfficiency}
A study found that using two alkali species, Cs and Li, for NEA activation could improve the chemical stability of GaAs photocathode under CO$_2$ exposure.\cite{mulhollan2008_EnhancedChemicalImmunity,sun2009_SurfaceActivationLayer,kurichiyanil2019test}
Recently, multiple research groups reported significantly improved lifetime using a few nm thick semiconductor activation layer with elements that form more robust photocathodes, such as Cs, Sb, Te, O.\cite{kuriki2019_NegativeElectronAffinity,biswas2021_HighQuantumEfficiency, rahman2019_NewActivationTechniques}
We reported a factor of 5 improvements in charge lifetime at 532 nm wavelength with Cs-Te activation layer\cite{bae2018_RuggedSpinpolarizedElectron} and a factor of 7 improvements at 780 nm with Cs-Sb-O layer.\cite{bae2020_ImprovedLifetimeHigh} Moreover, these layers showed a minimal adverse effect on spin-polarization.\cite{bae2018_RuggedSpinpolarizedElectron,bae2020_ImprovedLifetimeHigh,cultrera2020_LongLifetimePolarized}
However, these photoemission measurements were done with a 10's of eV bias and 100's of nA beam current, which is significantly different from the harsh environment of an electron gun that operates on the scale of 100's of keV. Specifically, the amount of current extracted and the energy-dependent residual gas ionization cross-sections are orders of magnitude different.\cite{grames2011_ChargeFluenceLifetime} Furthermore, the primary QE degradation mechanism was chemical poisoning in our previous works as opposed to ion back bombardment.\cite{cultrera2020_LongLifetimePolarized} In this work, we operated Cs-Sb-O and Cs-O activated GaAs photocathodes in the High ElectRon Average Current for Lifetime ExperimentS (HERACLES) beamline at Cornell University in order to make a direct comparison of their performance at 200 keV beam energy at 1 mA average current.
We saw a spectral dependence on lifetime improvement between the two types of samples and observed 45\% improvement in charge lifetime at 780 nm.

\section{Growth}

Highly \emph{p}-doped (Zn $5 \times 10^{18}$ cm$^{-3}$) GaAs (100) wafers were cleaved in air with a diamond scribe. Cleaved samples were rinsed in de-ionized water and solvent cleaned with isopropanol. We performed wet-etching with 1\% HF acid for 30 s, and then the samples were loaded into the vacuum for heat cleaning at $\sim 500$ C$^\circ$  for $\sim 48$ hours.
Cs-O activation was performed in a vacuum chamber directly connected to the back of the electron gun that has a pressure of $\sim 10^{-10}$ Torr. Resistively heated dispensers from SAES Getters were used for the Cs source, and O$_2$ was injected with a mechanical leak valve. We activated GaAs samples with the Cs-O$_2$ codeposition method at room temperature.

\begin{figure}
	\includegraphics*[width=230pt]{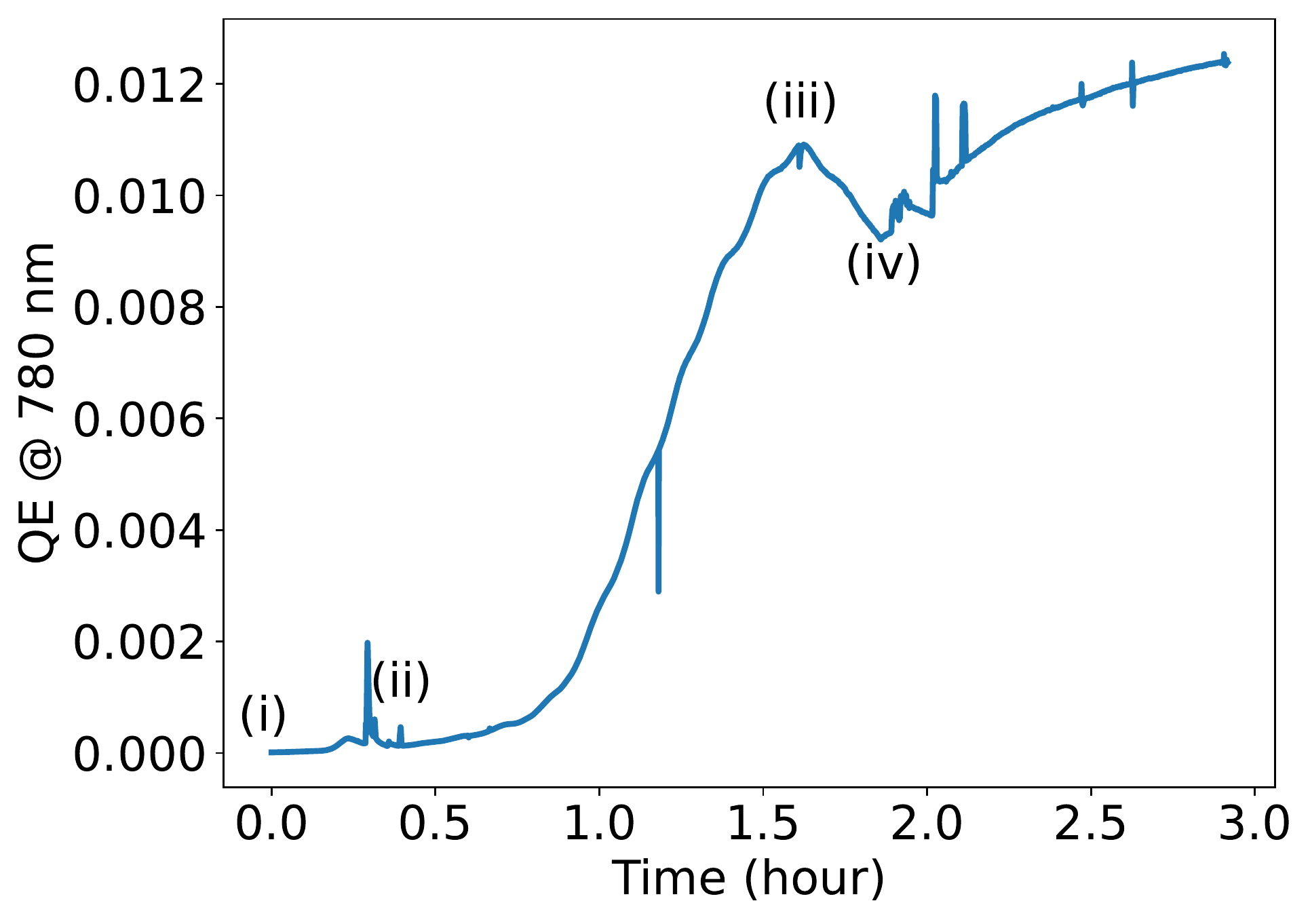}
	\caption{Quantum Efficiency at 780 nm of GaAs photocathode during Cs-Sb-O activation. Cs shutter was opened at (i), and the O$_2$ leak valve was opened at (ii). Sb shutter was opened at (iii) and closed at (iv).}
	\label{fig_growth22}
\end{figure}

For Cs-Sb-O activation, the growth was performed in a separate vacuum system located in another laboratory, and the cathodes were subsequently transported with a vacuum suitcase. The growth chamber has a pressure of $\sim 10^{-9}$ Torr, and single filament effusion cells were installed as Cs and Sb sources. Shutters in front of each source act as an on/off control of the flux. We leaked O$_2$ into the chamber with a mechanical leak valve similar to the Cs-O activation chamber. The temperature of the samples was kept at $\sim 130$ C$^\circ$ during the growth. QE at 780 nm was monitored during the growth with a diode laser ($\sim$ 10 $\mu$W) under 18 V negative bias.\cite{feng2019_activation}
Fig.~\ref{fig_growth22} demonstrates the QE at 780 nm during the growth process. (i) Samples were initially activated with just Cs until QE peaks and started decreasing. (ii) As soon as the QE decrease was observed, O$_2$ was leaked into the chamber. (iii) the Sb shutter was opened when QE saturated to a percent level. Cs, Sb, and O$_2$ are all being deposited simultaneously during Sb deposition. (iv) Sb shutter was closed after depositing 0.2 nm. The heater was turned off, the O$_2$ leak valve was closed, and the Cs shutter was closed soon after. The Sb deposition amount and flux ($5 \times 10^{11}$ atoms/cm$^2$/s) were estimated using a quartz crystal microbalance. The decreased QE during Sb deposition recovered while cooling down of the sample.
The activated cathodes were transported to the gun using a custom ultrahigh vacuum suitcase that has a base pressure of $< 10^{-10}$ Torr.\cite{galdi2020_EffectsOxygeninducedPhase} The suitcase was constantly pumped by a compact 400 l/s non-evaporative getter during the transport, and an 8 l/s ion pump was turned on once it was attached to the load-lock chamber of HERACLES. We typically pumped the load-lock chamber overnight before transporting the cathodes into the gun.


\section{Instrument}
\begin{figure*}
	\includegraphics*[width=400pt]{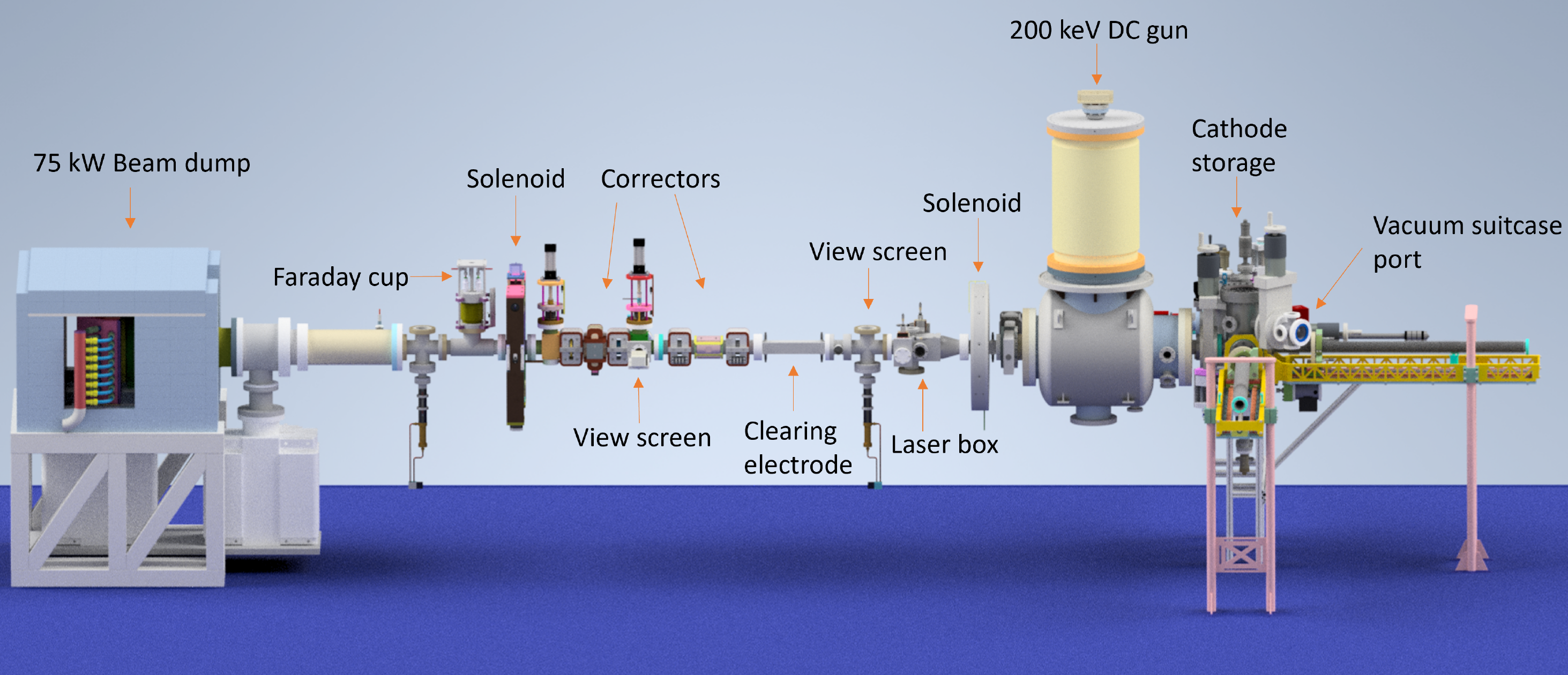}
	\caption{HERACLES beamline. Eletron beam travels from the gun to the beam dump. The beamline is consisted of two solenoids, horizontal/vertical corrector pairs, view screens, and a faraday cup.}
	\label{fig_heracles}
\end{figure*}

The HERACLES beamline consists of a 200 keV electron gun and a 75 kW beam dump (See Fig.~\ref{fig_heracles}). The gun was originally fabricated for the Cornell Electron Recovery Linac (ERL) program. It was used to demonstrate the record highest average current obtained from a photocathode.\cite{dunham2013_RecordHighaverageCurrent} Recently, the gun, beam dump, and some of the beamline components were moved to the same building as the Cornell Photocathode laboratory to enable an experimental program for producing and testing long-lived photocathodes at high average beam currents.
In the current configuration, beyond the gun, the HERACLES beamline comprises two solenoids, three horizontal/vertical corrector pairs, three BeO pneumatically mounted view screens, a faraday cup for low current ($\sim$100 nA) measurements, an emittance measuring system, and a quadrant detector for beam position monitoring at high currents. The beam dump is composed of 16 chilled water lines that are independently monitored with thermocouples that serve as a further diagnostic during high-current running. The beamline also includes two sets of clearing electrodes to reduce ion-back bombardment.

\begin{figure}
	\includegraphics*[width=230pt]{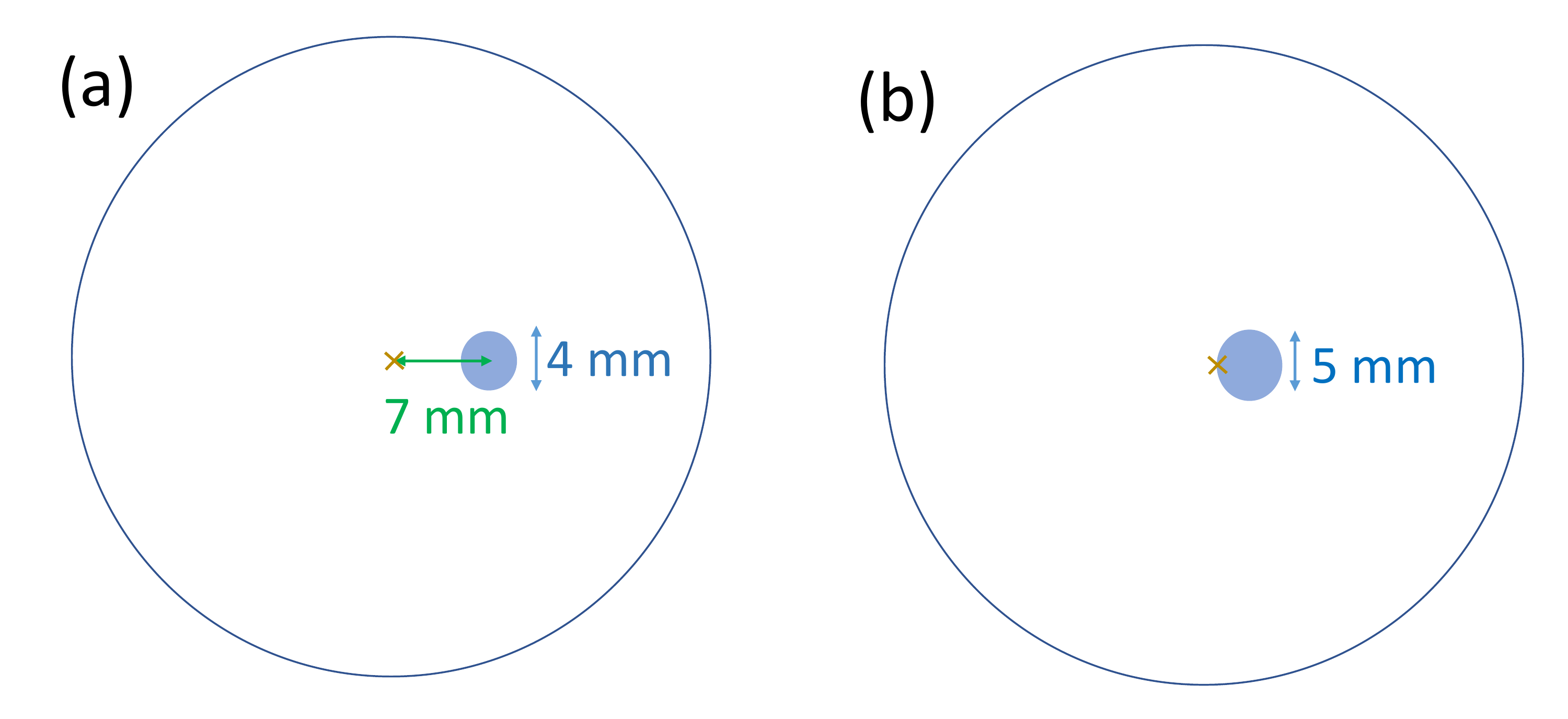}
	\caption{Active area (blue shaded area) geometry relative to the electrostatic center of the gun (yellow cross). (a) At Cornell DC electron gun, the active area is 7 mm off from the center. (b) At Jefferson Lab DC gun, the active area is closer to the electrostatic center.\cite{grames2011_ChargeFluenceLifetime}}
	\label{fig_puck22}
\end{figure}

During the high current runs presented in this work, we used Sabre Innova Argon gas laser by Coherent operated in the visible at 488 nm. A 1-to-1 imaging system consisting of a single 750 mm lens is used to image a 1.2 mm diameter pinhole onto the cathode active area. A small portion of the laser light is split off into a diagnostics optics path to allow the power to be monitored during the run.
The laser spot was positioned on the 7 mm off-centered active area on the cathode that has 4 mm diameter (see Fig.~\ref{fig_puck22} (a)).
Our control system is EPICS-based.\cite{dalesio1991epics} For running at 100 $\mu$A or more, the beam current is inferred from the current drawn by the gun power supply. During a charge lifetime measurement, the laser power is adjusted to maintain a constant current via a proportional-integral feedback loop implemented in Python with PYEPICS.

Two vacuum chambers are connected behind the HERACLES gun.
The first chamber has a mechanical arm for placing and removing cathodes in the gun, a vacuum attachment port for accepting cathodes via a vacuum suitcase, and has the capacity to store up to three cathodes. The second chamber, dubbed the QE-mapping chamber, has a cesium source and oxygen leak valve to enable Cs-O activation of GaAs. In this chamber, the saddle is electrically isolated and can be negatively biased to allow for photocurrent to be monitored during an activation. Outside this chamber, a diode laser is mounted to a 2D motorized stage to obtain transverse maps of the cathode's QE (Fig.~\ref{fig_map22}). Such a diagnostic is critical for an examination of the cathode after a high-current run to identify the main driver(s) of the cathode's performance degradation.

\section{Results and Discussion}

\begin{figure}
	\includegraphics*[width=230pt]{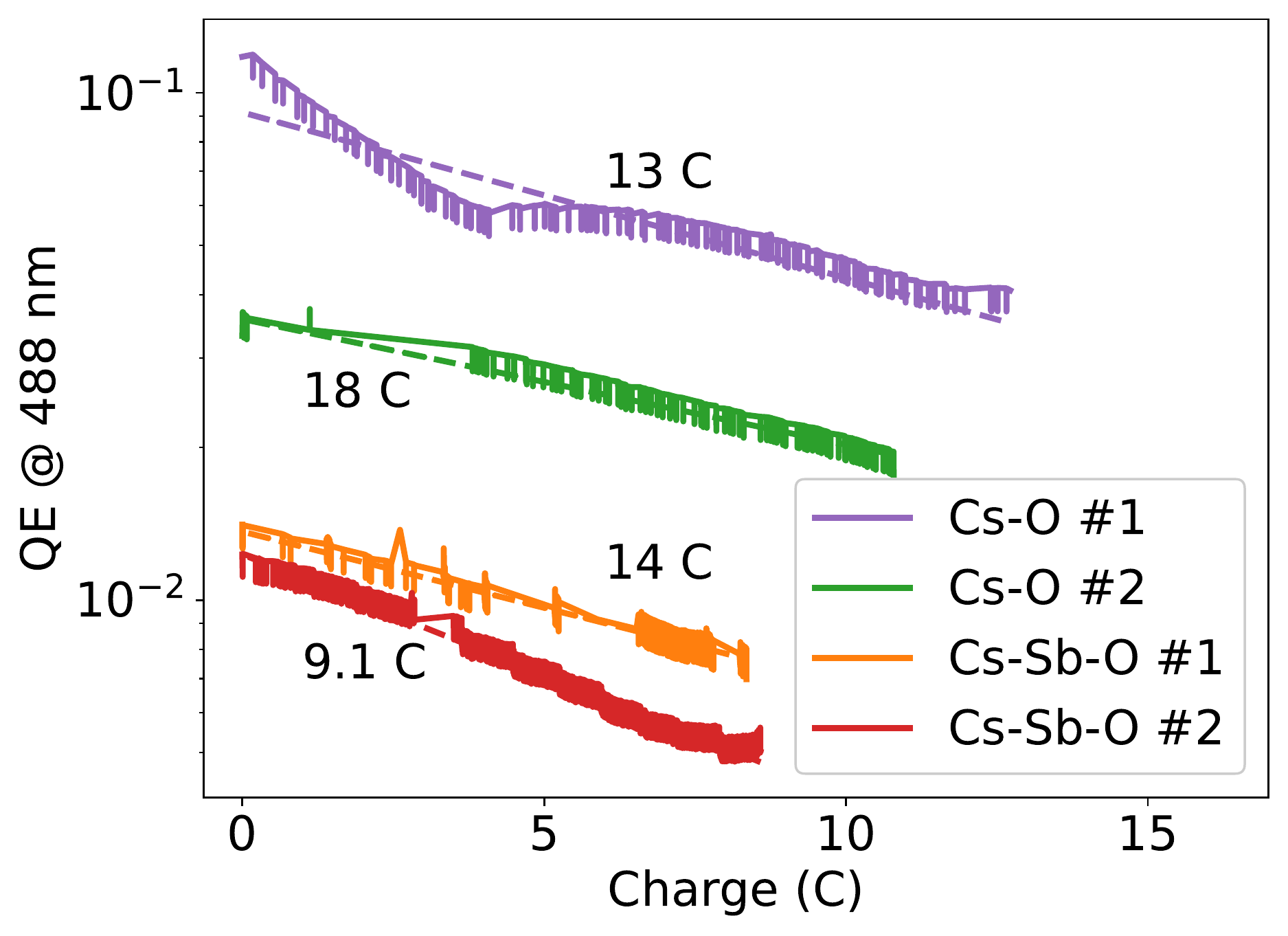}
	\caption{Quantum Efficiency degradation during 200 kV DC electron gun operation. 488 nm Laser was used to extract 1 mA of current. The solid lines represent measured data, while the dashed lines are exponential fits to the solid lines. The numbers next to each curve are estimated charge lifetimes from the exponential fit functions.}
	\label{fig_lifetime22}
\end{figure}

Charge lifetime is defined as the amount of charge extracted until QE drops to 1/\emph{e} of the initial value.\cite{grames2011_ChargeFluenceLifetime,bae2018_RuggedSpinpolarizedElectron}
In Fig.~\ref{fig_lifetime22}, QE at 488 nm was monitored during 200 keV operation at 1 mA. 
We obtained charge lifetimes around 15 C for the reference Cs-O activated GaAs photocathodes.
It must be noted that this value is two orders of magnitude lower than the ones achieved at the Thomas-Jefferson lab.\cite{grames2011_ChargeFluenceLifetime}
We believe there are two main reasons.
First, the pressure of the DC gun at Cornell is typically around $\sim 2 \times 10^{-11}$ Torr in a standby mode, and it can rise up to $10^{-10}$ Torr during an operation. On the other hand, the electron gun in the Jefferson Lab is operated under mid $10^{-12}$ Torr pressure.\cite{grames2011_ChargeFluenceLifetime} Secondly, it has been found that a high voltage photogun can produce unwanted photoelectrons from the edge of an active area that is not properly transported away.\cite{grames2011_ChargeFluenceLifetime,sinclair2007_DevelopmentHighAverage} These electrons follow extreme trajectories and strike the anode or beam pipe wall, resulting in vacuum degradation and, consequently, QE loss due to increased rates of vacuum poisoning and ion-bombardment. Jefferson lab showed improvements in charge lifetime by limiting the size of the active area to 5 mm and positioning the active area near the electrostatic center.\cite{grames2011_ChargeFluenceLifetime}
They achieved the largest charge lifetime when the laser spot was positioned at the furthest spot of the active area (5 mm).
A comparison of the active area geometry is demonstrated in Fig.~\ref{fig_puck22}. Although our active area has a similar size to that of the Jefferson Lab, the active area is far off from the center and can cause vacuum and QE degradation. When the active area has a 12.8 mm diameter size positioned near the electrostatic center, the Jefferson Lab was able to achieve only 10's of C for charge lifetime.\cite{grames2011_ChargeFluenceLifetime}
Considering the geometry, 15 C of charge lifetime agrees with measurements from the Jefferson Lab within an order of magnitude.
This agreement suggests that the HERACLES beamline is appropriate to perform comparison studies between the standard activated GaAs and Cs-Sb-O activated GaAs.

\begin{figure*}
	\includegraphics*[width=300pt]{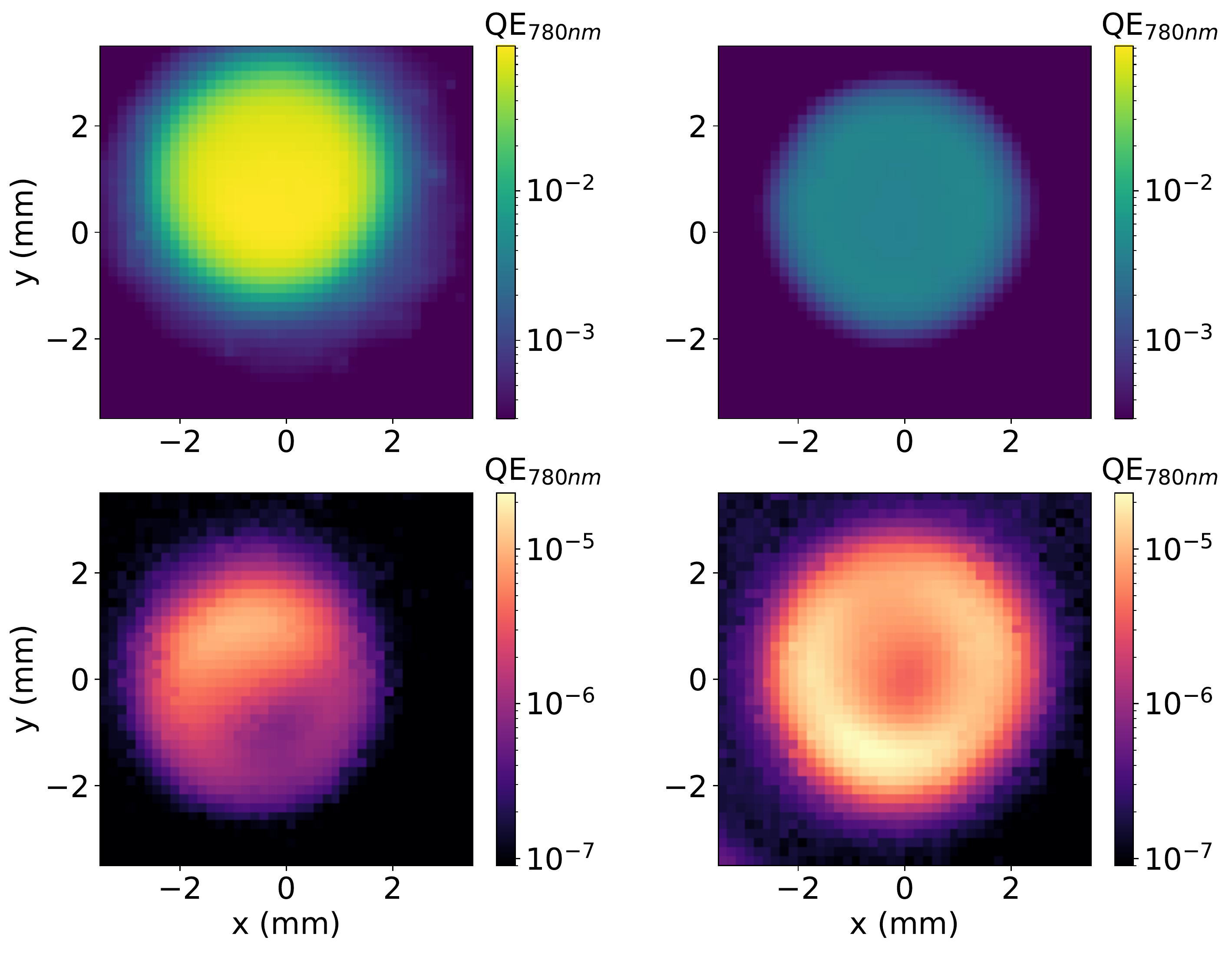}
	\caption{Quantum Efficiency map of active areas. 780 nm diode laser was used. The left images are Cs-O activated GaAs, while the right images are Cs-Sb-O activated GaAs. The top images were taken before the gun operation, and the bottom images were taken after.}
	\label{fig_map22}
\end{figure*}

In Fig.~\ref{fig_lifetime22}, we report the QE at 488 nm as a function of extracted charge measurements for two Cs-Sb-O activated samples and two Cs-O activated samples.
Regardless of the initial QE differences between the two types of activation layers, the charge lifetimes at 488 nm are within a factor of 2 of each other.
The initial QE of Cs-Sb-O activated samples are smaller compared to the standard Cs-O activated samples for two reasons. First, in our previous works, we observed roughly a factor 2-3 smaller QE when Sb was used during the activation procedure.\cite{cultrera2020_LongLifetimePolarized}
Secondly, the growths of Cs-Sb-O activated GaAs were done in a separate chamber, and the transport was done with a vacuum suitcase. This transport process takes about 24 hours, and we observed that the QE decreased by a factor of 3 after the transport was completed. On the other hand, the standard Cs-O activations are done in an attached chamber to the electron gun, and the samples were operated soon after the growths.
Because of the smaller QE, the average laser intensity during the Cs-Sb-O runs was necessarily higher, and, consequently, the rate of thermal desorption was faster. According to the thermal analysis found in reference \onlinecite{kuriki2011_DarklifetimeDegradationGaAs}, the temperature of the GaAs would have increased by a few Kelvin at most during the run. This same paper reports a few degree temperature rises have a negligible impact on the dark lifetime of GaAs activated with Cs-O. If we assume Cs-Sb-O to be equally robust to temperature, it is unlikely the lower QE impacted our lifetime measurements compared to ion back bombardment. However, thermal analysis is one of the topics for future works as it heavily depends on the geometry and materials.\cite{mammei2013_ChargeLifetimeMeasurements}

Postmortem QE maps were obtained with 780 nm light for the Cs-O activated sample \#1 and Cs-Sb-O activated sample \#2 in Fig.~\ref{fig_map22}.
The QE map demonstrates non-uniform damage on the surface, indicating that high current beam extraction caused the damage as opposed to vacuum poisoning, which we would expect to be uniform.
Despite one order of magnitude lower initial QE at 780 nm, we measured a higher final QE at 780 nm for Cs-Sb-O activated sample at the damaged area ($3.8 \times 10^{-6}$) compared to that of the Cs-O activated sample ($7.1 \times 10^{-7}$). This suggests that the Cs-Sb-O layer formed a more robust NEA layer in terms of chemical stability.
Although the charge lifetime at 780 nm was not measured directly, based on the amount of charge extracted during operation ($Q_{tot}$) and the initial and final QE (QE$_{i,f}$), we can estimate the charge lifetime at 780 nm ($\tau_{c}$) by assuming an exponential decay:
\begin{equation}
	\label{eq_lifetime}
	\begin{aligned}
		\text{QE}_{f} = \text{QE}_{i} \; e^{-Q_{tot}/\tau_{c}} \\
		\therefore \tau_{c} = \frac{Q_{tot}} {\log(\text{QE}_i/\text{QE}_f)}.
	\end{aligned}
\end{equation}
In Table ~\ref{table_lifetime}, the charge lifetimes at 780 nm are estimated for the four samples.
The estimated charge lifetime at 780 nm is 1.6 $\pm$ 0.4 C for Cs-Sb-O activated GaAs and 1.1 $\pm$ 0.1 C for Cs-O activated GaAs.
Our results show that the ratio of charge lifetimes of Cs-Sb-O and Cs-O is wavelength-dependent. While it is evident from Fig.~\ref{fig_lifetime22} that there is no improvement at 488 nm, we observe enhanced lifetimes for the former at 780 nm.

\setlength{\tabcolsep}{4pt}	
\begin{table*}
	\centering
	
	\begin{tabular}{l|llll}
		\hline \hline \\
		\multirow{2}{5em}{Activation method} & \multirow{2}{5em}{Initial QE at 780 nm}            & \multirow{2}{5em}{Final QE at 780 nm}               & 
		\multirow{2}{5em}{Extracted charge (C)} & \multirow{2}{5em}{Lifetime at 780 nm (C)} \\
		 \\
		\hline
		\\Cs-O$_2$ \#1 &  $9.1 \times 10^{-2}$  &  $ 7.1 \times 10^{-7}$ & $13$  &$1.1$       \\
		Cs-O$_2$ \#2 & $4.6 \times 10^{-2}$ & $2.1 \times 10^{-6}$	& $11$ &$ 1.1 $\\
		Cs-Sb-O \#1  & $4.2 \times 10^{-3}$     & $5.2 \times 10^{-5}$  &   $8.4$     & 1.9     \\
	
		Cs-Sb-O \#2	& $3.8 \times 10^{-3}$	& $3.8 \times 10^{-6}$	& $8.6 $ & 1.3	\\

		\hline \hline                                                                             
	\end{tabular}
	\caption{Estimated lifetimes at 780 nm based on Eq.~\ref{eq_lifetime}.}
	\label{table_lifetime}
\end{table*}

\begin{figure}
	\includegraphics*[width=230pt]{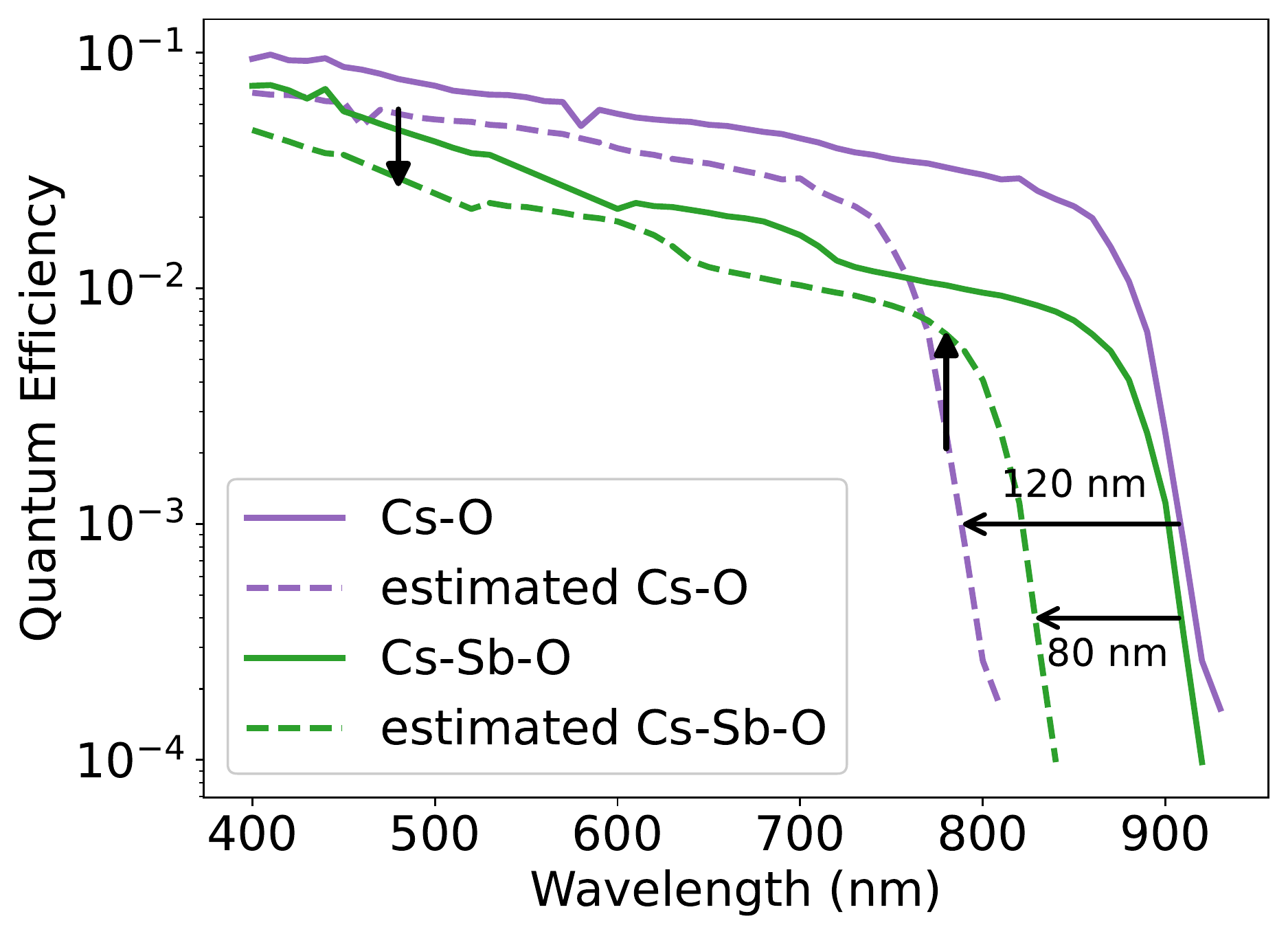}
	\caption{Estimated shifts in spectral response due to surface damage. The solid lines are typical spectral responses of Cs-O activated GaAs and Cs-Sb-O activated GaAs from our previous work.\cite{cultrera2020_LongLifetimePolarized} The dashed lines are estimated spectral responses of damaged cathodes. We assumed 5 C of charge is extracted for both samples. The spectral response shifts were 120 nm for Cs-O activated GaAs and 80 nm for Cs-Sb-O activated GaAs. The vertical black arrows indicate QE differences of damaged cathodes at 488 nm and 780 nm.}
	\label{fig_degrade22}
\end{figure}

This paragraph attempts to qualitatively explain the spectral dependence of lifetime improvement based on spectral response measurements we performed in previous work.\cite{cultrera2020_LongLifetimePolarized,bae2020_ImprovedLifetimeHigh}
Assuming the QE degradation is mainly due to damage done to the NEA layer, we can model the spectral response to simply shift horizontally during beam operation because of the increased work function.
Based on this assumption, estimated shifts in spectral response (QE$(\lambda)$) are plotted in Fig.~\ref{fig_degrade22}.
The solid lines are the spectral response curves of Cs-O activated GaAs and Cs-Sb-O activated GaAs from our previous measurements.\cite{cultrera2020_LongLifetimePolarized,bae2020_ImprovedLifetimeHigh}
The dashed lines have been obtained by simply shifting the same curves along the wavelength axis until the QE at 488 nm drops to the estimated QE after a 5 C extraction.
Note 5 C is about half of what we extracted in Fig.~\ref{fig_lifetime22}. The charge lifetimes at 488 nm in Fig.~\ref{fig_lifetime22} were used (15 C for Cs-O activated GaAs and 12 C for Cs-Sb-O activated GaAs).
It is estimated that the spectral response would shift 120 nm for Cs-O activated GaAs and 80 nm for Cs-Sb-O activated GaAs.
Despite a shorter lifetime at 488 nm for the Cs-Sb-O layer, the horizontal shift, or the work function increase, was smaller. This is because the log slope of spectral response ($d\log(\text{QE})/d\lambda$) at 488 nm was steeper for Cs-Sb-O activated sample ($6.4 \times 10^{-3}$/nm) compared to that of Cs-O activated sample ($2.8 \times 10^{-3}$/nm).
In other words, the sensitivity of the work function is the culprit for the shorter lifetime at 488 nm of the Cs-Sb-O activated sample, but the work function increase itself was smaller.
Since the work function increased less due to the robust NEA layer, the final QE at 780 nm was greater than that of the standard activated GaAs.




\section{Conclusion}

In this work, we operated Cs-O and Cs-Sb-O activated GaAs photocathodes at 200 keV with 1 mA average beam current to observe if a more chemically robust activation layer translated into an improvement in the charge lifetime under a high voltage environment.
We observed a 45\% improvement in charge lifetime at 780 nm for the Cs-Sb-O activated GaAs. This improvement at infrared wavelength does not correspond to an improvement at 488 nm, where instead, the lifetime was shorter. These results are interpreted in terms of work function increase induced by ion-back-bombardment. The larger lifetime at 780 nm suggests a smaller work function increase for the Cs-Sb-O case. Our results show that lifetime measurements at a single wavelength are not sufficient to fully characterize the robustness of NEA layers in operational conditions. 

For future work, characterizing spectral response during the beam operation can help understand the nature of spectral dependence of lifetime improvements. The effect of thermal desorption also needs to be identified.
Lastly, Cs-Sb-O activated GaAs can have a high initial QE by activating in a chamber connected to the beamline.

\section{Acknowledgments}

This work was supported by the Department of Energy under Grant No.DE-SC0021425 and the National Science Foundation under Grant No.PHY-1549132, the Center for Bright Beams.
The authors would like to acknowledge Joe Grames for valuable discussions.


\bibliographystyle{unsrt}
\bibliography{refch4, reference} 

\end{document}